\documentclass[aps,prd,english,preprintnumbers,
showpacs,showkeys,floatfix,nofootinbib,superscriptaddress,fleqn]{revtex4-1}
\usepackage{bm}
\usepackage{amstext}
\usepackage{amsmath,amsfonts}

\usepackage{babel}

\begin{document}

\preprint{INHA-NTG-09/2010}

\title{Electromagnetic mass differences of SU(3) baryons 
 within a chiral soliton model}

\author{Ghil-Seok Yang}

\email{Ghil-Seok.Yang@tp2.rub.de}

\affiliation{Institut f\"{u}r Theoretische Physik II,
  Ruhr-Universit\"{a}t Bochum, D--44780 Bochum, Germany}

\affiliation{Department of Physics, Inha University, Incheon 402-751, Republic
of Korea}

\author{Hyun-Chul Kim}

\email{hchkim@inha.ac.kr}

\affiliation{Department of Physics, Inha University, Incheon 402-751, Republic
of Korea}

\author{Maxim V. Polyakov}

\email{Maxim.Polyakov@tp2.rub.de}

\affiliation{Institut f\"{u}r Theoretische Physik II,
Ruhr-Universit\"{a}t Bochum, D--44780 Bochum, Germany}

\affiliation{Petersburg Nuclear Physics Institute, Gatchina, St. Petersburg 188
350, Russia}

\date{September, 2010}

\begin{abstract}
We investigate the electromagnetic mass differences of SU(3) baryons,
using an ``model-independent approach'' within a chiral soliton
model. The electromagnetic self-energy corrections to the masses of
the baryon are expressed as the baryonic two-point correlation function
of the electromagnetic currents. Using the fact that the electromagnetic
current can be treated as an octet operator, and considering possible
irreducible representations of the correlation function, we are able
to construct a general collective operator for the electromagnetic
self-energies, which consists of three unknown parameters. These
parameters are fixed, the empirical data for the electromagnetic mass
differences of the baryon octet being employed. We predict those of
the baryon decuplet and antidecuplet. In addition, we obtain various
mass relations between baryon masses within the corresponding
representation with isospin symmetry breaking considered. 
We also predict the physical mass differences of the baryon
decuplet. The results are in good agreement with the existing data.   
\end{abstract}

\pacs{12.39.Fe, 13.40.Em, 12.40.-y, 14.20.Dh}

\keywords{Electromagnetic mass differences of SU(3) baryons, Mass splittings
of SU(3) baryons, chiral soliton model}

\maketitle
\textbf{1.} Isospin symmetry breaking in mass splittings of hadrons 
has been one of the most fundamental issues historically even before
the advent of QCD~\cite{FeynmanSpeisman}. Effects of isospin symmetry
breaking are well known that they are originated from two different
sources: Electromagnetic (EM) self-energies (also known as photon
cloud energies) and the mass difference of the up and down quarks.
The electromagnetic (EM) self-energies contribute to masses of a hadron
isospin multiplet, depending on the corresponding charge. The EM mass
differences arising from the EM self-energies were extensively studied~
\cite{FeynmanSpeisman,Coleman:1964zz,Coleman:1961jn,Cottingham,GalScheck,
Nagylaki:1970tf,Elitzur:1970yy,Zee:1971df,Gunion}.
Gasser and Leutwyler analyzed in their seminal paper~\cite{Gasser:1982ap}
the EM contribution to baryon masses. They estimated the isospin-breaking
mass differences of the baryon octet with the EM self-energies and
were able to separate the pure hadronic part of isospin mass splittings,
which should arise from the up and down quark mass differences, subtracting
the EM mass differences from the experimental ones.

Thus, the mass splittings of the SU(3) baryons within an isospin multiplet
can fall into two different terms, i.e., the \textit{hadronic} and
\textit{electromagnetic} parts 
\begin{equation}
\Delta M_{B}\;=\; M_{B_{1}}-M_{B_{2}}\;=\;(\Delta
M_{B})_{\mathrm{H}}+(\Delta M_{B})_{\mathrm{EM}}, 
\label{eq:1}
\end{equation}
where the subscript $B$ denotes the baryon isospin multiplet and
$M_{B_{1}}$ and $M_{B_{2}}$ stand for masses of two different baryons
belonging to the same isospin multiplet. The $(\Delta M_{B})_{\mathrm{H}}$
and the $(\Delta M_{B})_{\mathrm{EM}}$ represent the hadronic and
the electromagnetic contributions to the mass splitting. There is
a great amount of works on the isospin mass differences of
baryons~\cite{Chemtob:1985ar,Li:1986iya,Jain:1989kn,Praszalowicz:1992gn,
Blotz:1994pc,Adami:1993xz,Dorokhov:1994fv}. 
Prasza{\l}owicz et al. calculated isospin mass splittings of the
baryon octet and decuplet~\cite{Praszalowicz:1992gn} in the chiral
quark-soliton model, considering the EM mass differences based on
the Dashen Ansatz~\cite{Dashen}. Though the analysis of
Ref.~\cite{Praszalowicz:1992gn} works well phenomenologically, the
Dashen Ansatz was originally derived for mesons and is valid strictly
in the chiral limit. 

While the isospin mass splittings for the baryon octet are
experimentally well known, those for the baryon decuplet are less
known. In the Review of Particle Physics 2010~\cite{Nakamura:2010pdg},
only the mass difference $M_{\Xi^{*0}}-M_{\Xi^{*-}}$ is given as
$-2.9\pm0.9$ MeV. The Gatchina group~\cite{Dakhno:1980cz} reported
many years ago the mass difference
$M_{\Delta^{++}}-M_{\Delta^{-}}=-5.9\pm3.1$ 
MeV. However, the experimental uncertainty is rather large.

In the present work, we want to investigate the EM mass differences
$(\Delta M_{B})_{\mathrm{EM}}$ of SU(3) baryons within the framework
of a chiral soliton model ($\chi$SM) in a {}``model-independent
approach''. The motivation lies in the fact that previous works on
the mass splittings of the SU(3) baryons in chiral soliton models
are hampered by some uncertainties~\cite{Diakonov:1997mm,Ellis:2004uz} 
in determining model parameters from the experimental data, since
they are rather sensitive to the data. One way to
discard these uncertainties is to turn on the isospin symmetry breaking
so that one can utilize the whole data of the masses of the baryon
octet. However, since the experimental data contain the effects of
isospin symmetry breaking due to both hadronic and EM self-energy
corrections, one has to extract the hadronic part from the data. In
order to isolate that, we have to subtract the EM self-energy
contributions from them so that we can fix unambiguously the model
parameters that are purely hadronic.

The EM self-energy corrections to the masses of the baryon are obtained
from the baryonic two-point correlation functions of the EM currents.
Thus, in this work, we will first derive a general collective operator
for the EM self-energies within a framework of the chiral soliton
model, using the fact that the EM current can be treated as an octet
operator, and considering possible irreducible representations of
the correlation functions. We will see that the collective operator
for the EM corrections consists of three unknown parameters. Instead
of computing these parameters using a model, we will fix them by employing
the EM mass differences of the baryon octet that were extracted in
Ref.~\cite{Gasser:1982ap}. With these parameters fixed, we can determine
not only the EM mass splittings of the SU(3) baryons but also the
\textit{physical} mass differences within the isospin multiplets. 
\\

\textbf{2.} The current quark mass term in the QCD Lagrangian can be
expressed as 
\begin{equation}
-\mathcal{L}_{\mathrm{m}}
\;=\;
\bar{\psi}\hat{m}\psi
\;=\;
 m_{\mathrm{u}}\bar{\mathrm{u}}\mathrm{u}
+m_{\mathrm{d}}\bar{\mathrm{d}}\mathrm{d}
+m_{\mathrm{s}}\bar{\mathrm{s}}\mathrm{s},
\label{eq:Lm}
\end{equation}
where the $\psi$ represents the quark field
$\psi=(\mathrm{u},\,\mathrm{d},\,\mathrm{s})$. The $\hat{m}$ denotes
the quark mass matrix $\hat{m}\;=\;\mathrm{diag}(m_{\mathrm{u}},\,
m_{\mathrm{d}},\, m_{\mathrm{s}})$. The Lagrangian can be expressed in
terms of the SU(3) flavor matrices 
\begin{equation}
-\mathcal{L}_{\mathrm{m}}\;=\;\bar{\psi}\left(m_{0} I\;+\;
  m_{3}\lambda_{3}\;+\; m_{8}\lambda_{8}\right)\psi 
\label{eq:lm}
\end{equation}
 with 
$m_{0}=(m_{\mathrm{u}}+m_{\mathrm{d}}+m_{\mathrm{s}})/3$,
$m_{3}=(m_{\mathrm{u}}-m_{\mathrm{d}})/2$, 
$m_{8}=(m_{\mathrm{u}}+m_{\mathrm{d}}-2m_{\mathrm{s}})/2\sqrt{3}$.
 The $I$ denotes the singlet flavor matrix $\mathrm{diag}(1,1,1)$
whereas $\lambda_{3}$ and $\lambda_{8}$ stand for the third
and the eighth  components of the flavor SU(3) Gell-Mann matrices,
respectively. Since the first term in Eq.(\ref{eq:lm}) is a flavor
singlet, it does not contribute to the mass splittings and mixings. On
the contrary, the second and the third ones in Eq.(\ref{eq:lm}), which
cause the isospin and SU(3) symmetry breakings, respectively, lead to
the mass splittings inside SU(3) baryon multiplets and the mixings
between the multiplets. They can be obtained by sandwiching the second
and third mass terms between the baryon states:  
$\langle B'|\bar{\psi}(m_{3}\lambda_{3}+m_{8}\lambda_{8})\psi|B\rangle$.

Taking into account the fact that $\lambda_{8}$
transforms as the eighth component of an octet operator, we can express
the masses of the baryon octet in terms of two parameters, since there
are two irreducible singlet representations from the direct product
of ${\bm{8}}\otimes\bm{8}\otimes\bm{8}$. On the other hand, the decuplet
and the antidecuplet are uniquely determined by a single parameter,
respectively, because the products ${\bm{8}}\otimes\bm{8}\otimes\bm{10}$
and ${\bm{8}}\otimes{\bm{8}}\otimes\overline{\bm{10}}$ contain only
one irreducible singlet representation, respectively. 
Similarly, we can parametrize the hadronic contributions of isospin
symmetry breaking as $(M_{8})_{H}^{\mathrm{iso}} = a\,
T_{3}+b\, Y\, T_{3}$, $(M_{10})_{H}^{\mathrm{iso}} = r\, T_{3}$, and 
$(M_{\overline{10}})_{H}^{\mathrm{iso}} = s\, T_{3}$,
where the $T_{3}$ and the $Y$ stand for the third component of
the isospin and the hypercharge, respectively. Thus, the hadronic
parts of the SU(3) baryon masses can be written as 
\begin{eqnarray}
M_{8} & = & \overline{M}_{8}\;+\;\left[
\begin{array}{c}
(a+b)T_{3}\\
0\\
aT_{3}\\
(a-b)T_{3}
\end{array}\right]\;+\;\left[
\begin{array}{c}
-3\, x\;+\; y\\
-x\\
x\\
2\, x\;-\; y
\end{array}\right]\;
\mbox{ for }\left[
\begin{array}{c}
N\\
\Lambda\\
\Sigma\\
\Xi
\end{array}\right],
\label{eq:8M2}\\
M_{10} & = & \overline{M}_{10}\;+\;\left[
\begin{array}{c}
rT_{3}\\
rT_{3}\\
rT_{3}\\
0
\end{array}\right]\;+\;\left[
\begin{array}{c}
-z\\
0\\
z\\
2z
\end{array}\right]\;\mbox{ for }\left[
\begin{array}{c}
\Delta\\
\Sigma^{*}\\
\Xi^{*}\\
\Omega^{-}
\end{array}\right],
\label{eq:10M2}\\
M_{\overline{10}} 
& = & 
\overline{M}_{\overline{10}}\;+\;\left[
\begin{array}{c}
0\\
sT_{3}\\
sT_{3}\\
sT_{3}
\end{array}\right]\;+\;\left[
\begin{array}{c}
-2v\\
-v\\
0\\
v
\end{array}\right]\;
\mbox{ for }\left[
\begin{array}{c}
\Theta^{+}\\
N^{*}\\
\Sigma_{\overline{10}}\\
\Xi_{3/2}
\end{array}\right],
\label{eq:10barM2}
\end{eqnarray}
where $M_{8}$, $M_{10}$, and $M_{\overline{10}}$ denote the mass
matrices corresponding to the baryon octet, decuplet, and antidecuplet,
respectively. The $\overline{M}_{8}$, $\overline{M}_{10}$, and
$\overline{M}_{\overline{10}}$ are the corresponding {\em center}
values. The parameters $x$, $y$, $z$, and $v$ are originated from the
flavor SU(3) symmetry breaking ($m_{8}$). The contributions from SU(3)
symmetry breaking are presented in the bases of
$(N,\,\Lambda,\,\Sigma,\,\Xi)$ for the octet,
$(\Delta,\,\Sigma^{*},\,\Xi^{*},\,\Omega^{-})$ for the decuplet, and
$(\Theta^{+},\, N^{*},\,\Sigma_{\overline{10}},\,\Xi_{3/2})$ for the 
antidecuplet. The isospin symmetric parts of Eq.~(\ref{eq:8M2})
are identified as the well-known Gell-Mann-Okubo mass formula~ 
\cite{GellMann:1962xb,Okubo:1961jc}.

 Note that, however, the baryon masses given in
 Eqs.(\ref{eq:8M2}-\ref{eq:10barM2}) do not contain the EM
 corrections. Thus, in order to get the \textit{physical} baryon
 masses, we have to take into account the EM corrections as 
in Eq.(\ref{eq:1}). We will show in the following how to analyze
these EM corrections to the baryon masses. 
\\

\textbf{3.} It is well known that the EM corrections to the baryon masses can
be derived from the baryonic two-point correlation functions of the
EM current $J_{\mu}$ in the static limit~\cite{Deshpande:1976vn}:
\begin{equation}
M_{B}^{\mathrm{EM}}\;=\;
\frac{1}{2}\int d^{3}x\, d^{3}y\langle
B|T[J_{\mu}(\bm{x})J^{\mu}(\bm{y})]|B\rangle D_{\gamma}(\bm{x},\bm{y}) 
\;=\;\langle B|\mathcal{O}^{\mathrm{EM}}|B\rangle,
\label{eq:corr}
\end{equation}
 where $J^{\mu}$ is defined as 
\begin{equation}
J^{\mu}(x)\;=\; e\bar{\psi}(x)\gamma_{\mu}\hat{Q}\psi(x)
\label{eq:currentJ}
\end{equation}
 with the electric charge $e$ and the quark charge operator $\hat{Q}$
defined as the Gell-Mann-Nishijima relation 
\begin{equation}
\hat{Q}\;=\;\frac{1}{2}\left(\lambda_{3}+\frac{1}{\sqrt{3}}\lambda_{8}\right).
\label{eq:gn}
\end{equation}
 The static photon propagator $D_{\gamma}$ is given as $1/4\pi|\bm{x}-\bm{y}|$,
but it will be absorbed in parameters we will fit to experimental
data.

The EM current is taken as an octet operator, so that we can write
in the most general form the $\mathcal{O}_{\mathrm{EM}}$ as a collective
operator 
\begin{eqnarray}
\mathcal{O}^{\mathrm{EM}} 
& = & 
\alpha_{1}\sum_{i=1}^{3}D_{Qi}^{(8)}D_{Qi}^{(8)}
+\alpha_{2}\sum_{p=4}^{7}D_{Qp}^{(8)}D_{Qp}^{(8)}
+\alpha_{3}D_{Q8}^{(8)}D_{Q8}^{(8)},
\label{eq:emop}
\end{eqnarray}
 where $D_{Qa}^{(8)}=(D_{3a}^{(8)}+D_{8a}^{(8)}/\sqrt{3})/2$ in which
$D_{ab}^{(8)}$ denote the SU(3) Wigner $D$ functions in the octet
representation. The parameters $\alpha_{i}$ encode specific dynamics
of a chiral soliton model. For example, one could use the chiral
quark-soliton model ($\chi$QSM) to obtain Eq.~(\ref{eq:emop}) and
determine $\alpha_{i}$. The EM operator $\mathcal{O}_{\mathrm{EM}}$
is expressed in the $\chi$QSM as 
\begin{equation}
\mathcal{O}_{\mathrm{EM}} \;=\; -\frac{e^{2}}{2}\int d^{3}x\,
d^{3}yD_{\gamma}(\bm{x},\,\bm{y})
\int\frac{d\omega}{2\pi}\mathrm{tr}\left\langle 
\bm{x}\left|\frac{1}{\omega+iH}\gamma_{\mu}\lambda^{a}
\right|\bm{y}\right\rangle \left\langle \bm{y}\left|
\frac{1}{\omega+iH}\gamma_{\mu}\lambda^{b}\right|\bm{x}\right\rangle 
D_{Qa}^{(8)}D_{Qb}^{(8)},
\label{eq:QEM}
\end{equation}
 where $D_{Qa}^{(8)}=(D_{3a}^{(8)}+D_{8a}^{(8)}/\sqrt{3})/2$. The
parameters $\alpha_{i}$ depends on specific dynamics of a $\chi$SM,
which will be fitted to the empirical data of the EM mass differences.
Since the EM current is regarded as an octet operator, the product
of two octet operators can be expressed in terms of irreducible operators
$\mathbf{1}\oplus\mathbf{8_{s}}\oplus\mathbf{8_{a}}
\oplus\mathbf{10}\oplus\mathbf{\overline{10}}\oplus\mathbf{27}$.
However, because of Bose symmetry, we are left only with the singlet,
the octet, and the eikosiheptaplet, which are all symmetric. A similar
structure for the EM corrections can be found in Ref.~\cite{deSwart:1963gc}.
We rewrite $\mathcal{O}^{\mathrm{EM}}$ in terms of a new set of parameters
$c^{(n)}$ as follows 

We can reduce the collective EM operator $\mathcal{O}_{\mathrm{EM}}$
in terms of single $D$ functions as follows 
\begin{equation}
\mathcal{O}^{\mathrm{EM}} 
\;=\; c^{(27)}\left(\sqrt{5}D_{\Sigma_{2}^{0}\Lambda_{27}}^{(27)}
+\sqrt{3}D_{\Sigma_{1}^{0}\Lambda_{27}}^{(27)}
+D_{\Lambda_{27}\Lambda_{27}}^{(27)}\right)\;+\;
\,c^{(8)}\left(\sqrt{3}D_{\Sigma^{0}\Lambda}^{(8)}
+D_{\Lambda\Lambda}^{(8)}\right)
+c^{(1)}D_{\Lambda\Lambda}^{(1)},
\label{eq:emop3}  
\end{equation}
where
\begin{equation}
  c^{(27)} \;=\; \frac{1}{40}\left(\alpha_{1}-4\alpha_{2} +
   3\alpha_{3}\right),\;\;\;\;\;
c^{(8)}\;=\; \frac{1}{10}\left(\alpha_{1}-\frac{2}{3}\alpha_{2} -
   \frac{1}{3}\alpha_{3}\right), \;\;\;\;\;
 c^{(1)}\;=\;\frac{1}{8}(\alpha_{1}+\frac{4}{3}\alpha_{2} +
 \frac{1}{3}\alpha_{3}).
\label{eq:c123}  
\end{equation}
The notations $\Sigma^{0}$, $\Sigma_{1}^{0}$, $\Sigma_{2}^{0}$, $\Lambda$,
and $\Lambda_{27}$ in the subscripts of the $D$ functions stand
for the corresponding flavor quantum numbers, in given representations
$\mathcal{R}$, $(Y,\, T,\, T_{3})^{(\mathcal{R})}=(0,\,1,\,0)^{(8)}$,
$(0,\,1,\,0)^{(27)}$ $(0,\,2,\,0)^{(27)}$, $(0,\,0,\,0)^{(8)}$,
and $(0,\,0,\,0)^{(27)}$, respectively~\cite{alfaro}. Note that
$\mathcal{O}_{\mathrm{EM}}$ in Eq.(\ref{eq:emop3}) consist only
of the eikosiheptaplet ($\bm{27}$), the octet ($\bm{8}$), and the
singlet ($\bm{1}$) representations. Since the EM current is regarded
as an octet operator, the product of two octet operators can be expressed
in terms of irreducible operators 
$\mathbf{1}\oplus\mathbf{8_{s}}\oplus\mathbf{8_{a}}\oplus
\mathbf{10}\oplus\mathbf{\overline{10}}\oplus\mathbf{27}$.
However, because of Bose symmetry, we are left only with the singlet,
the octet, and the eikosiheptaplet, which are all symmetric. 
Note that the first three terms of Eq.~(\ref{eq:emop3}) (part of the
eikosiheptaplet) have the same parameter $c^{(27)}$ of the
contributions from the $\Delta T\,=\,0,\,1,\,2$
transitions in eikosiheptaplet. The last term in Eq.(\ref{eq:emop3})
does not contribute to the EM mass splittings, because it is the singlet
and corresponding corrections will be canceled out for the EM mass
differences. The parameters $c^{(n)}$ will be fixed by the empirical
data estimated in Ref.~\cite{Gasser:1982ap}. A similar
structure for the EM corrections can be found in
Ref.~\cite{deSwart:1963gc} in which the operator for the EM mass
splitting is given as 
\begin{equation}
\mathcal{O}_{\mathrm{Swart}}^{EM} \;=\; \mathcal{O}_{(0,2,0)}^{(27)} +
 \mathcal{O}_{(0,1,0)}^{(27)} + \mathcal{O}_{(0,1,0)}^{(8)} +
 \mathcal{O}_{(0,0,0)}^{(1)}.  
\label{eq:deswart}
\end{equation} 
The operator~(\ref{eq:deswart}) is different from Eq.(\ref{eq:emop3}),
because $\Delta T=1$ and $\Delta T=2$ contributions in the
eikosiheptaplet are treated separately in Eq.~(\ref{eq:deswart}). On
the other hand, they are identical in the EM operator of the present
work. This is due to the fact that in the chiral soliton model
the SU(2) soliton is embedded in SU(3) by Witten's trivial
embedding~\cite{Witten:1983tw}.  

In order to calculate the matrix elements of Eq.(\ref{eq:emop3}),
we need to know SU(3) baryon wave functions. In the $\chi$SM, the
baryon wave functions are found to be SU(3) Wigner functions in
representation $\mathcal{R}$ 
\begin{equation}
|B\rangle\;=\;\sqrt{\mathrm{dim}(\mathcal{R})}(-1)^{J_{3}+Y'/2}
D_{(Y,T,T_{3})(-Y',J,J_{3})}^{(\mathcal{R})*}(A),
\label{eq:ketB}
\end{equation}
 which diagonalize the collective Hamiltonian in the $\chi$SM. The
$Y'$ denotes the eighth component of the SU(3) spin operator 
$Y'=-2J_{8}/\sqrt{3}=N_{c}B/3=1$. The $B$ is the baryon number. The
constraint of the $Y'$ in the Skyrme model arises from the Wess-Zumino
term~\cite{Witten:1983tx,Guadagnini:1983uv} 
whereas it comes from the valence quarks filled in the discrete level
in the $\chi$QSM~\cite{Blotz:1992pw,Christov:1995vm}. The collective
baryon wave functions are not in a pure representation, when the SU(3)
symmetry breaking effects are considered. However, since we are interested
in the EM mass differences in the present work, we need not consider
the wave-function corrections. 

The EM mass can be obtained by sandwiching the collective operator
$\mathcal{O}_{\mathrm{EM}}$ in Eq.(\ref{eq:emop}) between the baryon
states. The corresponding results can be written for the baryon octet
%%%%%%%% EM masses for the baryon octet %%%%%%%%%%%%%%%%%%%%%%%
\begin{eqnarray}
M_{N}^{\mathrm{EM}} & = & 
\frac{1}{5}\left(c^{(8)}+\frac{4}{9}c^{(27)}\right)T_{3}+\frac{3}{5}\left(c^{(8)}
+\frac{2}{27}c^{(27)}\right)\left(T_{3}^{2}+\frac{1}{4}\right)+c^{(1)},\cr 
M_{\Lambda}^{\mathrm{EM}} & = & 
\frac{1}{10}\left(c^{(8)}-\frac{2}{3}c^{(27)}\right)+c^{(1)},\cr
M_{\Sigma}^{\mathrm{EM}} & = & 
\frac{1}{2}c^{(8)}\, T_{3}+\frac{2}{9}c^{(27)}\, T_{3}^{2}
-\frac{1}{10}\left(c^{(8)}+\frac{14}{9}c^{(27)}\right)+c^{(1)},\cr
M_{\Xi}^{\mathrm{EM}} & = & 
\frac{4}{5}\left(c^{(8)}-\frac{1}{9}c^{(27)}\right)T_{3}-\frac{2}{5}\left(c^{(8)}
-\frac{1}{9}c^{(27)}\right)\left(T_{3}^{2}+\frac{1}{4}\right)+c^{(1)},
\label{eq:emforoc}
\end{eqnarray}
 and for the baryon decuplet 
\begin{eqnarray}
M_{\Delta}^{\mathrm{EM}} & = & 
\frac{1}{4}\left(c^{(8)}+\frac{8}{63}c^{(27)}\right)T_{3}
+\frac{5}{63}c^{(27)}\, T_{3}^{2}+\frac{1}{8}\left(c^{(8)}-\frac{2}{3}c^{(27)}\right)+c^{(1)},\cr
M_{\Sigma^{\ast}}^{\mathrm{EM}} & = & 
\frac{1}{4}\left(c^{(8)}-\frac{4}{21}c^{(27)}\right)T_{3}
+\frac{5}{63}c^{(27)}\,\left(T_{3}^{2}-1\right)+c^{(1)},\cr
M_{\Xi^{\ast}}^{\mathrm{EM}} & = & 
\frac{1}{4}\left(c^{(8)}-\frac{32}{63}c^{(27)}\right)T_{3}
-\frac{1}{4}\left(c^{(8)}+\frac{8}{63}c^{(27)}\right)\left(T_{3}^{2}+\frac{1}{4}\right)+c^{(1)},\cr
M_{\Omega^{-}}^{\mathrm{EM}} & = &
 -\frac{1}{4}\left(c^{(8)}-\frac{4}{21}c^{(27)}\right)+c^{(1)},
\label{eq:emfordec}
\end{eqnarray}
 respectively. Since the center of baryon masses can absorb the singlet
contributions to the EM masses with $c^{(1)}$, we can safely neglect
them for EM mass differences. Moreover, they are not pertinent to
the EM mass differences in which they are canceled out. Therefore,
the expressions of EM mass differences of SU(3) baryons have only
two unknown parameters, i.e. $c^{(8)}$ and $c^{(27)}$. 

As shown in Eqs.(\ref{eq:emforoc}, \ref{eq:emfordec}), they are
expressed in terms of the isospin third component $T_{3}$, its square
$T_{3}^{2}$, and the constant terms arising from the hypercharge.
Note that Eqs.(\ref{eq:emforoc}, \ref{eq:emfordec}) in general can
be rewritten in terms of the electric charge $Q$ and its square $Q^{2}$
with the Gell-Mann-Nishijima relation in Eq.(\ref{eq:gn}) used. We
want to mention that the present results are distinguished from the
Dashen ansatz for the EM mass splittings that shows $Q^{2}$
proportionality ($\sim Q_{B}^{2}M_{B}$), which was
employed in Ref.~\cite{Praszalowicz:1992gn}. Moreover, it turns out that
Eqs.(\ref{eq:emforoc}, \ref{eq:emfordec}) have the same structures as the
Weinberg-Treiman mass formula $M(T_{3})\;=\;\alpha T_{3}^{2}+\beta
T_{3}+\gamma$~\cite{Weinberg:1959zzb}. 

It is straightforward to obtain the EM mass differences for the baryon
octet from Eq.(\ref{eq:emforoc}) 
\begin{equation}
(M_{p}-M_{n})_{\mathrm{EM}} \;=\;
\frac{1}{5}\left(c^{(8)}+\frac{4}{9}c^{(27)}\right), \;\;\;\;\;
(M_{\Sigma^{+}}-M_{\Sigma^{-}})_{\mathrm{EM}} \;=\; c^{(8)}, \;\;\;\;\;
(M_{\Xi^{0}}-M_{\Xi^{-}})_{\mathrm{EM}} \;=\;
\frac{4}{5}\left(c^{(8)}-\frac{1}{9}c^{(27)}\right).
\label{eq:diff_EM}  
\end{equation}
Using Eq.(\ref{eq:diff_EM}), we immediately obtain the following mass
formula 
$ c^{(8)}=(M_{p}-M_{n})_{\mathrm{EM}}+(M_{\Xi^{0}}-M_{\Xi^{-}})_{\mathrm{EM}}
= (M_{\Sigma^{+}}-M_{\Sigma^{-}})_{\mathrm{EM}}$.
This is just the well-known Coleman-Glashow mass formula~\cite{Coleman:1961jn}
at the level of the EM corrections. Although these formulae indicate
that these three mass differences are dependent on each other,
one can adjust the values of the parameters $c^{(8)}$ and $c^{(27)}$
by the method of least squares. In order to determine the parameters
$c^{(8)}$ and $c^{(27)}$, we will first use the empirical data
estimated in Ref.~\cite{Gasser:1982ap}. Using these empirical and
experimental data, we can determine the values of the parameters
$c^{(8)}$ and $c^{(27)}$ as follows  
\begin{eqnarray}
c^{(8)}\;=\;-0.15\pm0.23, 
&  & c^{(27)}\;=\;8.62\pm2.39
\label{eq:c1c2}
\end{eqnarray}
 in units of MeV. In Table~\ref{tab:Em_mass.diff}, the reproduced
EM mass differences for the baryon octet are listed. %
\begin{table}[ht]
\caption{EM mass differences for the baryon octet in units of MeV.}

\begin{tabular}{lcr}
\hline \hline
 & Inputs  & Reproduced \tabularnewline
\hline 
$\left(M_{p}-M_{n}\right)_{\mathrm{EM}}$  
& $\;\;0.76\pm0.30$ ~\cite{Gasser:1982ap}  
& $\;\;0.74\pm0.22$\tabularnewline
$\left(M_{\Sigma^{+}}-M_{\Sigma^{-}}\right)_{\mathrm{EM}}$  
& $-0.17\pm0.30$ ~\cite{Gasser:1982ap}  
& $-0.15\pm0.23$\tabularnewline
$\left(M_{\Xi^{0}}-M_{\Xi^{-}}\right)_{\mathrm{EM}}$  
& $-0.86\pm0.30$ ~\cite{Gasser:1982ap}  
& $-0.88\pm0.28$\tabularnewline
\hline \hline
\end{tabular}\label{tab:Em_mass.diff} 
\end{table}

Employing the results of Eqs.(\ref{eq:8M2}) and (\ref{eq:c1c2}) and
the masses of the baryon octet as input, we can determine the
parameters for the hadronic isospin symmetry breaking as follows  
\begin{equation}
a \;=\; -3.63\pm0.09,\;\;\; b\;=\;2.86\pm0.12
\label{eq:abxyM8}
\end{equation}
in units of MeV. From Eq.(\ref{eq:8M2}), we are able to reproduce
various mass relations of the baryon octet, the EM corrections being
taken into account  
\begin{eqnarray}
M_{p}\;-\; M_{n} & = & 
\left(M_{\Sigma^{+}}\;-\; M_{\Sigma^{-}}\right)\;-\;\left(M_{\Xi^{0}}-M_{\Xi^{-}}\right).
\label{eq:coleman}
\end{eqnarray}
Equation~(\ref{eq:coleman}) is the Coleman-Glashow
relation~\cite{Coleman:1961jn}. Note that even though we consider the
EM corrections, the Coleman-Glashow relation is still 
preserved. Taking into account both mass splittings from the SU(3) and
isospin symmetry breakings with the EM corrections, we derive the 
following relations  
\begin{eqnarray}
2\left(M_{p}+M_{\Xi^{0}}\right) 
& = & 3M_{\Lambda}+\overline{M}_{\Sigma}
+\left(M_{\Sigma^{+}}-M_{\Sigma^{-}}\right)+\frac{2}{3}\Delta M_{\Sigma},
\cr
2\left(M_{n}+M_{\Xi^{-}}\right) 
& = & 3M_{\Lambda}+\overline{M}_{\Sigma}
-\left(M_{\Sigma^{+}}-M_{\Sigma^{-}}\right)+\frac{2}{3}\Delta M_{\Sigma},
\label{eq:isoGO}
\end{eqnarray}
where $\Delta M_{\Sigma}\;=\; M_{\Sigma^{+}}+M_{\Sigma^{-}}-2M_{\Sigma^{0}}$. 
These two mass relations are well satisfied with the experiment
data. If we turn off the effects of isospin symmetry breaking, the
mass formulae of Eq.(\ref{eq:isoGO}) are reduced to the following
relation 
\begin{eqnarray}
2\left(\overline{M}_{N}+\overline{M}_{\Xi}\right) 
& = & 3M_{\Lambda}+\overline{M}_{\Sigma}+\frac{2}{3}\Delta M_{\Sigma},
\label{eq:SU3GO}
\end{eqnarray}
which generalizes Gell-Mann-Okubo mass formula for the baryon octet~
\cite{GellMann:1962xb,Okubo:1961jc}
with the EM corrections presented in the last term. When the EM
interaction is turned off, Eq.~(\ref{eq:SU3GO}) leads to the
Gell-Mann-Okubo mass formula  
\begin{eqnarray}
2\left(\overline{M}_{N}\;+\;\overline{M}_{\Xi}\right) & = &
3M_{\Lambda}\;+\;\overline{M}_{\Sigma}. 
\label{eq:GO}
\end{eqnarray}

The EM mass differences of the baryon decuplet can be read off from
Eq.~(\ref{eq:emfordec}) as follows 
%%%%%%%%%% EM mass differences for the baryon decuplet %%%%%%%%%%%
\begin{eqnarray}
\left(M_{\Delta^{++}}-M_{\Delta^{+}}\right)_{\mathrm{EM}} 
& = & \frac{1}{4}\left(c^{(8)}+\frac{16}{21}c^{(27)}\right),\cr
\left(M_{\Delta^{+}}-M_{\Delta^{0}}\right)_{\mathrm{EM}} 
&=& \left(M_{\Sigma^{\ast+}}-M_{\Sigma^{\ast0}}\right)_{\mathrm{EM}}
=\frac{1}{4}\left(c^{(8)}+\frac{8}{63}c^{(27)}\right),\cr
\left(M_{\Delta^{0}}-M_{\Delta^{-}}\right)_{\mathrm{EM}} 
& = & \left(M_{\Sigma^{\ast0}}-M_{\Sigma^{\ast-}}\right)_{\mathrm{EM}}
\;=\;\left(M_{\Xi^{\ast0}}-M_{\Xi^{\ast-}}\right)_{\mathrm{EM}} \;=\;
\frac{1}{4}\left(c^{(8)}-\frac{32}{63}c^{(27)}\right), 
\label{eq:em10diff}
\end{eqnarray}
for $\Delta T_{3}\;=\;1$, 
\begin{eqnarray}
\left(M_{\Delta^{++}}-M_{\Delta^{0}}\right)_{\mathrm{EM}} 
& = & \frac{1}{2}\left(c^{(8)}+\frac{4}{9}c^{(27)}\right),\cr
\left(M_{\Delta^{+}}-M_{\Delta^{-}}\right)_{\mathrm{EM}} 
& = & \left(M_{\Sigma^{\ast+}}-M_{\Sigma^{\ast-}}\right)_{\mathrm{EM}}
\;=\;\frac{1}{2}\left(c^{(8)}-\frac{4}{21}c^{(27)}\right),
\label{eq:em10diff2}
\end{eqnarray}
for $\Delta T_{3}\;=\;2$, and
\begin{eqnarray}
\left(M_{\Delta^{++}}-M_{\Delta^{-}}\right)_{\mathrm{EM}} 
& = & \frac{3}{4}\left(c^{(8)}+\frac{8}{63}c^{(27)}\right),
\label{eq:em10diff3}
\end{eqnarray}
for $\Delta T_{3}\;=\;3$.
The parameter $r$ in Eq.(\ref{eq:10M2}) is found to be 
\begin{equation}
r \;=\; -2.19\;\pm\;0.08
\label{eq:randz}
\end{equation}
in units of MeV. Note that the mass relations shown in
Eqs.(\ref{eq:em10diff})-(\ref{eq:em10diff3}) are also valid with the
hadronic effects of the isospin symmetry breaking considered. 
\\

\textbf{4.} We now present the numerical results and discuss them. Since we have
determined all relevant parameters for the EM mass differences of
the SU(3) baryons , i.e. $c^{(8)}$ and $c^{(27)}$,
we can proceed to calculate numerically the EM mass differences. Putting
into Eq.(\ref{eq:em10diff}) the numerical values of $c^{(8)}$ and
$c^{(27)}$ in Eq.(\ref{eq:c1c2}), we can predict the results of
the EM mass differences of the baryon decuplet. In Table~\ref{tab:dec1em},
we list the corresponding results. %
\begin{table}[t]
\centering 
\caption{EM mass differences of the baryon decuplet in units of MeV.}
\begin{tabular}{lc|cc}
\hline \hline
$(\Delta M_{B_{10}})_{\mathrm{EM}}$  
& Numerical results  
& $(\Delta M_{B_{10}})_{\mathrm{EM}}$  
& Numerical results \tabularnewline
\hline 
$(M_{\Delta^{++}}-M_{\Delta^{+}})_{\mathrm{EM}}$  
& $\;\;1.60\pm0.46$ 
& $(M_{\Delta^{++}}-M_{\Delta^{0}})_{\mathrm{EM}}$  
& $\;\;1.84\pm0.54$\tabularnewline
$(M_{\Delta^{+}}-M_{\Delta^{0}})_{\mathrm{EM}}$  
& $\;\;0.24\pm0.10$ 
& $(M_{\Delta^{+}}-M_{\Delta^{-}})_{\mathrm{EM}}$  
& $-0.89\pm0.26$\tabularnewline
$(M_{\Delta^{0}}-M_{\Delta^{-}})_{\mathrm{EM}}$  
& $-1.13\pm0.30$ 
& $(M_{\Delta^{++}}-M_{\Delta^{-}})_{\mathrm{EM}}$ 
 & $\;\;0.71\pm0.29$\tabularnewline
$(M_{\Sigma^{\ast+}}-M_{\Sigma^{\ast0}})_{\mathrm{EM}}$  
& $\;\;0.24\pm0.10$ 
& $(M_{\Sigma^{\ast+}}-M_{\Sigma^{\ast-}})_{\mathrm{EM}}$  
& $-0.89\pm0.26$\tabularnewline
$(M_{\Sigma^{\ast0}}-M_{\Sigma^{\ast-}})_{\mathrm{EM}}$  
& $-1.13\pm0.30$ 
&  & \tabularnewline
$(M_{\Xi^{\ast0}}-M_{\Xi^{\ast-}})_{\mathrm{EM}}$ 
& $-1.13\pm0.30$ 
&  & \tabularnewline
\hline\hline
\end{tabular}
\label{tab:dec1em} 
\end{table}
\begin{table}[h]
\caption{Isospin mass splittings of the baryon decuplet in units of MeV.}
\begin{tabular}{ccc}
\hline \hline
$(\Delta M_{B_{10}})$  & This work  & Experimental data \tabularnewline
\hline 
$(M_{\Delta^{++}}-M_{\Delta^{+}})$  
& $-0.59\pm0.47$ & \tabularnewline
$(M_{\Delta^{+}}-M_{\Delta^{0}})$  
& $-1.95\pm0.13$ & \tabularnewline
$(M_{\Delta^{0}}-M_{\Delta^{-}})$  
& $-3.32\pm0.32$ & \tabularnewline
$(M_{\Sigma^{\ast+}}-M_{\Sigma^{\ast0}})$  
& $-1.95\pm0.13$ & \tabularnewline
$(M_{\Sigma^{\ast0}}-M_{\Sigma^{\ast-}})$  
& $-3.32\pm0.32$ & $-3.1\pm0.6$~\cite{Thomas:1973uh}\tabularnewline
$(M_{\Xi^{\ast0}}-M_{\Xi^{\ast-}})$  
& $-3.32\pm0.32$ & $-2.9\pm0.9$~\cite{Nakamura:2010pdg}\tabularnewline
$(M_{\Delta^{++}}-M_{\Delta^{0}})$  
& $-2.54\pm0.57$ & $-2.86\pm0.30$~\cite{gridnevetal}\tabularnewline
$(M_{\Delta^{+}}-M_{\Delta^{-}})$  
& $-5.28\pm0.30$ & \tabularnewline
$(M_{\Delta^{++}}-M_{\Delta^{-}})$  
& $-5.86\pm0.38$ & $-5.9\pm3.1$ \cite{Dakhno:1980cz}\tabularnewline
$(M_{\Sigma^{\ast+}}-M_{\Sigma^{\ast-}})$  
& $-5.28\pm0.30$ & \tabularnewline
\hline\hline
\end{tabular}\label{tab:physdec} 
\end{table}

Since we have also determined the mass parameter $r$ for the hadronic
isospin symmetry breaking, we can easily find the numerical results
for the physical mass differences within the isospin multiplets of the
baryon decuplet, which are listed in Table~\ref{tab:physdec}. We
compare the present results with the experimental data for the isospin
mass splittings for the baryon decuplet. Note that the data from
Ref.~\cite{gridnevetal} are based on the $\pi N$ phase-shift
analysis. Moreover, one has to keep in mind that the experimental
data~\cite{Dakhno:1980cz} suffer from a large amount of errors and are
not completely free from model dependence. The present results are in
good agreement with the experimental data within uncertainties.
The results in Table~\ref{tab:physdec} consist of the contributions
from both the mass difference between the up and down quarks
($m_{\mathrm{u}}-m_{\mathrm{d}}$) and the EM corrections. Since we are
able to extract the EM corrections from the physical mass differences,
we can estimate how large isospin symmetry is hadronically broken. 
Moreover, we find that the \textit{physical} mass relation 
\begin{equation}
(M_{\Sigma^{\ast0}}-M_{\Sigma^{\ast-}}) \;=\;
(M_{\Xi^{\ast0}}-M_{\Xi^{\ast-}})    
\end{equation}
is well satisfied with the experimental data, as shown in
Table~\ref{tab:physdec}. 

For completeness, we also present the results of the EM mass
differences for the baryon antidecuplet listed in Table~\ref{tab:antidec}.
Though there are no experimental
data for them, it is still of great importance to know them in order
to determine the masses of the baryon antidecuplet unambiguously.
\begin{table}[b]
\caption{Electromagnetic mass differences of the baryon anti-decuplet in units
of MeV.}

\begin{tabular}{cc|cc}
\hline \hline
$(\Delta M_{B_{10}})_{\mathrm{EM}}$  
& This work  
& $(\Delta M_{B_{10}})_{\mathrm{EM}}$  
& This work \tabularnewline
\hline 
$\left(M_{p^{\ast}}-M_{n^{\ast}}\right)_{\mathrm{EM}}$  
& $-1.31\pm0.31$ 
& $\left(M_{\Sigma_{\overline{10}}^{+}}-M_{\Sigma_{\overline{10}}^{-}}\right)_{\mathrm{EM}}$ 
& $-0.89\pm0.26$\tabularnewline
$\left(M_{\Sigma_{\overline{10}}^{+}}-M_{\Sigma_{\overline{10}}^{0}}\right)_{\mathrm{EM}}$  
& $-1.31\pm0.31$ 
& $\left(M_{\Xi_{3/2}^{+}}-M_{\Xi_{3/2}^{-}}\right)_{\mathrm{EM}}$  
& $-0.89\pm0.26$\tabularnewline
$\left(M_{\Sigma_{\overline{10}}^{0}}-M_{\Sigma_{\overline{10}}^{-}}\right)_{\mathrm{EM}}$  
& $\;\;0.24\pm0.10$ 
& $\left(M_{\Xi_{3/2}^{0}}-M_{\Xi_{3/2}^{--}}\right)_{\mathrm{EM}}$  
& $-1.84\pm0.54$\tabularnewline
$\left(M_{\Xi_{3/2}^{+}}-M_{\Xi_{3/2}^{0}}\right)_{\mathrm{EM}}$  
& $-1.31\pm0.31$ 
& $\left(M_{\Xi_{3/2}^{+}}-M_{\Xi_{3/2}^{--}}\right)_{\mathrm{EM}}$  
& $\;\;0.71\pm0.29$\tabularnewline
$\left(M_{\Xi_{3/2}^{0}}-M_{\Xi_{3/2}^{-}}\right)_{\mathrm{EM}}$  
& $\;\;0.24\pm0.10$ 
&  & \tabularnewline
$\left(M_{\Xi_{3/2}^{-}}-M_{\Xi_{3/2}^{--}}\right)_{\mathrm{EM}}$ 
& $\;\;1.60\pm0.46$ 
&  & \tabularnewline
\hline\hline
\end{tabular}\label{tab:antidec} 
\end{table}
\\

\textbf{5.} In the present work, we have investigated the electromagnetic mass
differences of the SU(3) baryons, employing an ``model-independent
approach'' within a chiral soliton model. The electromagnetic self-energy
corrections to the masses of the baryon are expressed as the baryonic
two-point correlation function of the electromagnetic currents. We
first derived a general collective operator for the electromagnetic
self-energies, using the fact that the electromagnetic current can
be treated as an octet operator, and considering possible irreducible
representations of the correlation function. The collective operator
for the electromagnetic corrections were shown to have three unknown
parameters. Instead of computing them using a model, we
have fixed $c^{(8)}$ and $c^{(27)}$ by the empirical
data~\cite{Gasser:1982ap} for the electromagnetic 
mass differences of the baryon octet. The parameters $c^{(8)}$ and
$c^{(27)}$, which are responsible for the mass splittings of the
isospin multiplet due to the electromagnetic self-energies, were found
to be $-0.15\pm0.23$ MeV and $8.62\pm2.39$ MeV, respectively. 

Having calculated the electromagnetic mass differences, we were able
to extract the hadronic part of the isospin mass splittings from the
physical mass differences. The results of the \textit{physical}
isospin mass differences of the baryon decuplet are in good agreement
with the existing data. The mass relations are also well satisfied
with the data. In addition, we also present the
electromagnetic mass differences of the baryon antidecuplet for
completeness.    

Since we have determined the EM mass differences for the SU(3) baryons,
we can continue to study the mass splittings of the SU(3) baryons
unambiguously. The corresponding investigation will appear elsewhere.

%%%%%%%%%%%%%%%%%%%%%%%%%%%%%%%
\section*{Acknowledgments}
%%%%%%%%%%%%%%%%%%%%%%%%%%%%%%%
The present work is 
supported by Basic Science Research Program through the National
Research Foundation of Korea (NRF) funded by the Ministry of
Education, Science and Technology (grant number: 2010-0016265). 
The work is also supported by the Transregio-Sonderforschungsbereich
Bonn-Bochum-Giessen, the Verbundforschung (Hadrons and Nuclei) of the
Federal Ministry for Education and Research (BMBF) of Germany, the
Graduiertenkolleg Bochum-Dortmund, the COSY-project J{\"u}lich as well
as the EU Integrated Infrastructure Initiative Hadron Physics Project
under contract number RII3-CT-2004-506078.

\end{document}